\documentclass[showpacs,preprintnumbers,amsmath,amssymb,preprint]{revtex4}
\usepackage{amsmath,amssymb,amsfonts,latexsym}
\begin{document}
\title{Supergroup BF action for supergravity}
\author{C. Ram\'{\i}rez}
\email{cramirez@fcfm.buap.mx}
\affiliation{Facultad de Ciencias F\'{\i}sico Matem\'aticas, Universidad Aut\'onoma de
Puebla, P.O. Box 1364, 72000 Puebla, M\'exico.}
\author{E. Rosales}
\email{jerq15@hotmail.com}
\affiliation{Facultad de Ciencias F\'{\i}sico Matem\'aticas, Universidad Aut\'onoma de
Puebla, P.O. Box 1364, 72000 Puebla, M\'exico.}

\begin{abstract}
General relativity can be formulated as a complex BF-theory with constraints, as given by Pleba\'nski. We show that a straightforward $OSp(2|1)$ extension of the action requires additional fields with one constraint, and that supergravity with cosmological constant turns out on-shell, after the application of integrability conditions in the action. The action is invariant under Kalb-Ramond transformations which, as in usual BF theories, correspond on-shell to diffeomorphisms.
\end{abstract}
\pacs{4.65.+e, 4.60.-m, 4.50.Kd}
\maketitle

\section{Introduction}
\label{intro}
The problem of quantization of gravity is one the most important unsolved problems in physics. The most promising approaches to solve it have been superstring theory and loop quantization \cite{loop}. In the loop approach, quantization can be done in a diffeomorphism invariant setting, by means of a graph approach of the path integral formulation of Pleba\'nski complex BF action \cite{reisenberger1}, i.e. by spin-foam \cite{baez} or state-sum models, in which the metricity (simplicity) constraints of the $B$-field impose conditions on the allowed representations of the associated symmetry group \cite{pietri}. Inversely, starting from spin-foam models, the Pleba\'nski action turns out in the continuum limit \cite{crane,baez}.

In this search one important element has been given by gauge formulations. Such formulations have been useful in particular for the incorporation of supersymmetry with gravity to get supergravity \cite{nieuwenhuizen,wess}. Following this philosophy, MacDowell and Mansouri have written a gauge theory for gravity and for supergravity \cite{MM}, where the tetrad and the spin connection are components of a gauge field of the group $SO(3,2)$, respectively of the supergroup $Osp(1|4)$. In this approach, the action contains terms which explicitly break symmetry. Simultaneously to this work, Pleba\'nski has worked out a complex $BF$-action \cite{Plebanski}, with gauge invariance under $SL(2,C)$, which can be rewritten as a local Lorentz invariant action with self-dual fields \cite{jacobson}. The relation to gravity arises through the so called simplicity or metricity constraints, i.e. the traceless bilinear form of two $B$-fields vanish, whose solution is given in terms of a tetrad \cite{Plebanski}. Thus from a chiral, right handed constrained BF-theory, a non-chiral theory arises. These constraints can be included into the action by means of suitable Lagrange multipliers. From this action written in terms of the tetrad, the self-dual Palatini action follows, and thus the Ashtekar action as shown by Samuel \cite{samuel}. For a thorough analysis see e.g. \cite{capovilla,Krasnov}. A supergravity version, with an action invariant under $SL(2,C)$, has been given by Jacobson \cite{jacobson}. This action has been further considered by Capovilla et. al. \cite{capovilla}, who found suitable fermionic constraints acting on a two-form valued right handed spinor, and whose solution is given in terms of a one-form valued left-handed spinor. Invariant actions under $OSp(2|1)$ have been given by various authors \cite{sano,obregon,ezawa,nojiri,smolinsugra}, as well as a self-dual version of the MacDowell-Mansouri supergravity action \cite{nieto1}, with $OSp(4|1)$ gauge fields. $OSp(2|1)$ loop quantum supergravity has been also considered \cite{obregon,smolinsugra}, by means of extensions of the Ashtekar formulation. 
The BF action of Capovilla et.al. \cite{capovilla} with a cosmological constant term has been studied for $N=1$ and $N=2$ by various authors \cite{sano,ezawa,nojiri}. In these works, the BF part of the action has been written in an $Osp(2|1)$ covariant way \cite{ezawa}, and the constraints have been written as supertraces under certain irreducible representations of $OSp(2|1)$ \cite{nojiri}; they show also that the Kalb-Ramond on-shell symmetry \cite{kalb} breaks down, unless certain conditions on the vector transformation parameters are imposed. 
None of these actions has been obtained straightforwardly from the corresponding bosonic action, i.e. by substituting the fields by $Osp(2|1)$ fields and the trace by a supertrace, and the problem turns out to be that the simplicity constraints cannot be generalized directly, by a supertraceless bilinear form.

The renormalizability properties of the Pleba\'nski action have been studied in \cite{KrasnovR}, where it is argued that the quantum effects can be resumed by an effective, field dependent cosmological constant, leading then to a modified ``non-metric" formulation. Consequences of this theory have been studied for the Ashtekar formulation, for spin foams in \cite{KrasnovSF}, and in cosmology \cite{KrasnovC}. Another generalization is given in \cite{KrasnovD}, where by relaxing the constraints a generalized potential for the BF theory is proposed, and it is argued that it leads as well to general relativity.

Generalizations of the Pleba\'nski action, with real fields, have been studied in \cite{KrasnovHD}, where it is extended to higher dimensions, and in \cite{montesinos}. Recently, generalizations have been considered for extended groups in \cite{smolin1}, where a Yang-Mills sector arises from the extra gauge fields. More recently, the Pleba\'nski action has been traced back to a Matrix theory \cite{smolin2}. 

In \cite{KrasnovD}, Krasnov considered the Pleba\'nski action with the cosmological constant generalized by a field, and it turns out that the consistency of the equations of motion plus the simplicity constraints require that this field is constant. Thus this action leads on-shell to the action of Pleba\'nski. In \cite{montesinos2} an action with the same form, i.e. with a field instead the cosmological constant, but with real fields, has been related to a proposal by Husain and Kuchar \cite{kuchar}.

A straightforward supergroup generalization of the Pleba\'nski action would be expected to be given by the vanishing of the supertrace of the product of two $B$-fields. However, as we show, it leads to a nilpotent volume element. In fact, this seems to be the reason why in previous works, the supersymmetric simplicity constraints have not been written in a compact, covariant $OSp(2|1)$ form. In the search of this generalization we were lead to the action of Krasnov mentioned in the previous paragraph, with an additional constraint term which becomes relevant for the supergroup action \cite{Ramirez:2009zz}. Further we show that the equations of motion of the supergroup action plus the resulting generalized simplicity constraints, lead to similar consistency conditions as in the bosonic case, in such a way that when applied to the action, it follows Townsend supergravity \cite{townsend}. Furthermore for both actions, bosonic and supersymmetric, the Kalb-Ramond invariance can be implemented without extra conditions, and lead as usual to diffeomorphisms. In Sec. 2, we consider the $SU(2)$ theory. In Sec. 3 we consider the $OSp(2|1)$ generalization and in Sec. 4 we draw some conclusions.

\section{BF action of gravitation}
\label{sec:1}
Pleba\'nski has given a complex theory in \cite{Plebanski}. This theory can be formulated as a purely self-dual $SO(1,3)$ theory as shown by Jacobson \cite{jacobson}, who has shown also that this action amounts to a real theory because its imaginary part becomes a total derivative when the field equations of the Lorentz connection are considered. Let us take a $SU(2)$ complex one-form connection $ \Omega= \Omega_it^i$ $(i=1,2,3)$, with its two-form field strength $F=d \Omega+ \Omega\wedge\Omega$. The generators of the algebra satisfy $[t^i,t^j]=2i{\epsilon^{ij}}_kt^k$ and ${\rm Tr}(t^it^j)=2\delta^{ij}$, where $\epsilon_{ijk}$ is the Levi-Civita symbol. 

The Pleba\'nski action with cosmological constant $\Lambda$ is given by,
\begin{equation}
I=\frac{2i}{k}\int \left[{\rm Tr}B\wedge F+B^i\wedge\left(\Phi_{\ \ i}^{(t)\,j}+\frac{\Lambda}{3}\delta_i^{\ j}\right)B_j\right],\label{pleba0}
\end{equation}
where $B=B^it_i$ is a Lie algebra valued two-form, and the symmetric traceless tensor $\Phi_{ij}^{(t)}$ is a zero-form. The variation of $\Phi_{ij}^{(t)}$ leads to the simplicity constraints
\begin{equation}\label{constraints}
B^i\wedge  B^j-\frac{1}{3}\delta^{ij}B^k\wedge  B_k=0.
\end{equation}
This action can be straightforwardly generalized to be invariant under the supergroup $OSp(2|1)$, however this generalization leads to an inconsistent volume element, as will be seen in the next section. 
In \cite{KrasnovD} a generalization of this action has been pursued and as a first instance, the tensor field is analyzed by means of an expansion in the complex two-forms $B^i$ and $\bar B^i$, which are supposed to be compatible with the $SU(2)$ connection, i.e. $DB=0$. Hence $F^i(\Omega)={\cal F}^{ij}B_j+\bar {\cal F}^{ij}\bar B_j$. Then under the assumption of self-duality the second term of this equation is discarded. Further, considering the Bianchi identities satisfied by the tensor field, $DF=0$, then $D({\cal F}^{ij}B_j)=D({\cal F}^{ij})B_j=0$, which is shown to lead to $D_{\mu}({\cal F}^{ij})B_i\wedge B_j=0$, see later in this section. Now if the constraints (\ref{constraints}) are taken into account and supposing that $B^iB_i\neq 0$, the last equation turns into $\partial_\mu{\rm Tr}{\cal F}=0$. Hence the traceless part of ${\cal F}$ turns to be the Weyl tensor, and its trace leads to the the Einstein equation with cosmological constant, which in this case is written as ${\rm Tr}{\cal F}=\Lambda$. In \cite{KrasnovD} it is raised the question of a general action from which this setting follows. Our interest in this paper is not on this problem, but on an action which could be extended in a straightforward way to a supergroup action. As we argue in the next section, the following action does this job,
\begin{equation}
I=\frac{2i}{k}{\rm Tr}\int \left[B\wedge F+B\wedge\Phi(B)+\eta\Phi\right],\label{pleba}
\end{equation}
where $\Phi$ is an automorphism in the space of the Lie algebra valued two-forms, i.e. it is represented by a matrix, whose matrix elements are fields, which we write as follows 
Tr$B\wedge\Phi(B)=B^i\,\Phi_i^{\ j}B_j$.
Further, $\eta$ is a four form with values in the adjoint representation of $SU(2)$, hence 
\begin{equation}
{\rm Tr}(\eta\Phi)=\eta^i\epsilon_{ij}^{\ \ k}\Phi_k^{\ j}.
\end{equation}
This term is irrelevant at this stage because it eliminates the antisymmetric part of the matrix $\Phi$, and we will not consider it in the rest of this section.

Action (\ref{pleba}) satisfies in a specific way the above mentioned requirements. Indeed, in order to variate the independent degrees of freedom, $\Phi$ must be decomposed into its irreducible components, its trace $\phi={\rm Tr}\,\Phi$ and its traceless part
$\phi^{(t)}_{ij}=\Phi_{ij}-\frac{1}{3}\phi\delta_{ij}$, hence
\begin{equation}
I=\frac{2i}{k}\int \left(2 B^i\wedge F_i+\phi^{(t)}_{ij} B^i\wedge  B^j+
\frac{1}{3}\phi B^i\wedge  B_i\right).\label{pleba1}
\end{equation}

Further, a variation with respect to $\phi^{(t)}$ gives the constraints (\ref{constraints}), and the equations of motion corresponding to the variation of the fields $B$ and $\Omega$ are,
\begin{equation}\label{eqbf}
F_i+\Phi_{ij}B^j=0 \ {\rm and }\ DB^i=0.
\end{equation}
Considering the Bianchi identity $DF=0$, these equations lead to the consistency conditions
\begin{equation}\label{consistencia}
D\Phi_{ij}\wedge B^j=0,
\end{equation}
which suitably contracted with a $B$-field give
\begin{equation}\label{consistencia2}
\epsilon^{\mu\nu\rho\sigma}D_\nu\Phi_{ij}B_{\rho\sigma}^iB_{\mu\tau}^j=0,
\end{equation}
from which it follows that (see Appendix) \cite{KrasnovD},
\begin{equation}\label{consistencia3}
D_\mu\Phi_{ij}\epsilon^{\nu\rho\sigma\tau}B_{\nu\rho}^iB_{\sigma\tau}^j=0.
\end{equation}
Hence, taking into account the constraints (\ref{constraints}) we can write $D_\mu \Phi_{ij}\delta^{ij}=\partial_\mu \phi=0$, i.e. $\phi$ is constant and setting it into the action (\ref{pleba}) we recover (\ref{pleba0}). 

BF actions have, besides the obvious gauge symmetry, in our case $SU(2)$, a vector symmetry \cite{kalb,horowitz}, which when realized on-shell with field dependent parameters corresponds to diffeomorphism invariance \cite{horowitz}. For instance the BF action with `cosmological constant'
\begin{equation}
I=\int \left({\rm Tr}B\wedge F+\frac{1}{2}\Lambda B^i\wedge B_i\right),\label{bfl}
\end{equation}
is invariant under the $SU(2)$ transformations,
\begin{equation}\label{gauge}
\delta_\alpha B=-[\alpha,B]\quad {\rm and}\quad \delta_\alpha \Omega=d\alpha-[\alpha,\Omega],
\end{equation}
and the transformations,
\begin{equation}\label{kr0}
\delta_C B=-DC, \quad {\rm and}\quad \delta_C \Omega=\Lambda C,
\end{equation}
where $C$ are one-form transformation parameters, valued in the Lie algebra of $SU(2)$. It is well known \cite{horowitz} that the diffeomorphisms plus field dependent gauge transformations follow on-shell if the transformation parameters have the field dependent form 
\begin{equation}
C=-\mathnormal{i}_vB \quad {\rm i.e.}\quad C_\mu^i=v^\nu B_{\mu\nu}^{\ \ i},\label{sustitucion}
\end{equation}
consistently with the fact that from construction the action has diffeomorphism and gauge invariance. Action (\ref{pleba}) inherits these symmetries, although the Kalb-Ramond transformations can be realized only in the non-linear setting of (\ref{sustitucion}), and the transformations are now
\begin{eqnarray}
\delta_v B_{\mu\nu}^i&=&-D_\mu(v^\rho B_{\nu\rho}^{\ \ i})+D_\nu(v^\rho B_{\mu\rho}^{\ \ i}),\label{kr1}\\
\delta_v \Omega_\mu^i&=&2v^\nu\Phi^i_{\ j}B_{\mu\nu}^{\ \ j},\label{kr2}\\ 
\delta_v\Phi_{ij}&=&v^\mu D_\mu\Phi_{ij}.\label{kr3}\\
\delta_v\eta^i&=&D_\mu(v^\mu\eta^i).\label{kr4}
\end{eqnarray}
Indeed, if we apply these transformations to action (\ref{pleba}), by means of the Bianchi identitity $DF=0$ and ignoring total derivatives,  considering (\ref{sustitucion}) we get
\begin{equation}
\delta I=\frac{2i}{k}\int\left(2D\Phi_{ij}B^iC^j+v^\mu D_\mu\Phi_{ij}B^iB^j\right)=0.
\end{equation}
where the last equality is due to the identity (see appendix),
\begin{eqnarray}
D\Phi_{ij}B^iC^j&=&-\frac{1}{2}d^4x\epsilon^{\mu\nu\rho\sigma}D_\mu\Phi_{ij}B^i_{\nu\rho}B_{\sigma\lambda}^jv^\lambda\nonumber\\
&=&\frac{1}{8}d^4xv^\lambda D_\lambda\Phi_{ij}\epsilon^{\mu\nu\rho\sigma}B^i_{\mu\nu}B_{\rho\sigma}^j\nonumber\\
&=&-\frac{1}{2}v^\mu D_\mu\Phi_{ij}B^iB^j.
\end{eqnarray}
Further, on-shell, under use of the equations of motion $F^i+2\Phi^i_{\ j}B^j=0$ and $DB^i=0$, transformations (\ref{kr1})-(\ref{kr4}) become diffeomorphisms plus field dependent gauge transformations,
\begin{eqnarray}
\delta_v B_{\mu\nu}^i&=&v^\rho\partial_\rho B_{\mu\nu}^i+\partial_\mu v^\rho B^i_{\rho\nu}+
\partial_\nu v^\rho B^i_{\mu\rho}-\delta_\alpha B_{\mu\nu}^i,\quad\label{diff1}\\
\delta_v \Omega_\mu^i&=&v^\nu\partial_\nu \Omega_\mu^i+\partial_\mu v^\nu \Omega_\nu^i
-\delta_\alpha \Omega_\mu^i,\label{diff2}\\
\delta_v\Phi_{ij}&=&v^\mu\partial_\mu\Phi_{ij}-\delta_\alpha\Phi_{ij},\label{diff3}\\
\delta_v\eta^i&=&\partial_\mu(v^\mu\eta^i)-\delta_\alpha\eta^i.\label{diff4}
\end{eqnarray}
where $\delta_\alpha$ are $SU(2)$ transformations with parameters $\alpha^i=v^\mu \Omega_\mu^i$. The four-forms $\eta^i$ transform as invariant densities.

Before we consider the supergroup action, we review how is obtained gravity from (\ref{pleba0}). The fields are decomposed into their real and imaginary parts,
$ \Omega=\frac{1}{2}(\omega+i\tilde\omega)$ and
$B=\frac{1}{2}(\Sigma+i\tilde\Sigma)$. Thus we get
$F=\frac{1}{2}(R+i\tilde{R})$, where
\begin{eqnarray}
R^i&=&d\omega^i-{\epsilon^i}_{jk}\omega^j\wedge \tilde\omega^k,\label{ri}\\
{\tilde{R}}^i &=&d\tilde\omega^i+\frac{1}{2}{\epsilon^i}_{jk}
(\omega^j\wedge\omega^k-\tilde\omega^j\wedge\tilde\omega^k).
\end{eqnarray}
Further, we define $\omega^{ij}={\epsilon^{ij}}_k{\tilde\omega}^k$
and $R^{ij}={\epsilon^{ij}}_k{\tilde R}^k$. Then,
\begin{equation}
R^{ij}=d\omega^{ij}+\omega^{il}\wedge\omega_l^{\ j}+\omega^i\wedge\omega^j.\label{rij}
\end{equation}
Thus, if $\omega^{i0}=-\omega^{0i}=-\omega^i$, $\omega^{00}=0$ and
$R^{0i}=R^i$, it follows that
\begin{equation}
R^{ab}=d\omega^{ab}+\omega^{ac}\wedge\omega_{c}^{\ b}\qquad (a,b=0,1,2,3).\label{R}
\end{equation}
The Minkowski metrics signature is $\eta^{ab}={\rm diag}(-1,1,1,1)$.
In a similar way, a two tensor $\Sigma^{ab}$ can be obtained from $B^i$. Thus, 
we get the self-dual quantities 
\begin{equation}\label{descomp}
\Omega^i=\omega^{(+)0i}=
\frac{1}{2}\left(\omega^{0i}-\frac{i}{2}\epsilon^{0i}_{\ \ cd}\omega^{cd}\right)
\end{equation}
as well as $B^i=\Sigma^{(+)0i}$ and $R^i=R^{(+)0i}$,
which satisfy $R^{(+)}(\omega)=R(\omega^{(+)})$. These self-dual quantities satisfy 
$\epsilon^{ab}_{\ \ cd}\Sigma^{(+)cd}=2i\Sigma^{(+)ab}$.

Thus, the BF part of the action turns into a self-dual action,
\begin{equation}\label{accionsd}
I_{BF}=-\frac{i}{k}\int \Sigma^{(+)ab}\wedge R^{(+)}_{ab}.
\end{equation}

AS a well known fact, the solution of the constraints (\ref{constraints})
is given by a factorization of $\Sigma^{ab}$ as the product of two tetrad real one-forms \cite{Plebanski}, and 
give a basis in the space of the two-forms, 
\begin{equation}\label{solconstr}
\Sigma^{ab}=e^a \wedge e^b.
\end{equation}

Thus, considering these results and (\ref{accionsd}), the action (\ref{pleba1}) can be written as,
\begin{equation}
I=-\frac{1}{4k}\int\bigg(
\frac{1}{2}\epsilon_{abcd}\Sigma^{ab}\wedge\Sigma^{ef}R_{ef}^{\ \ cd}\,d^4x
+\frac{\Lambda}{6}\epsilon_{abcd}\Sigma^{ab}\wedge\Sigma^{cd}
+2i\Sigma^{ab}\wedge R_{ab}\bigg).\label{12}
\end{equation}
Further, by means of the Bianchi identity $DT^a= R^a_{\ b}\wedge e^b$, where $T^a=De^a$ is the torsion, this action can be written as the sum of the Palatini action with cosmological constant, plus a term linear in the torsion
\begin{equation}
I=-\frac{1}{2k}\int [R(\omega)+2\Lambda]ed^4x-
\frac{i}{4k}\int e^a\wedge DT_a,\label{accionsd2}
\end{equation}
where $e={\rm Det}(e_\mu^{\ a})$.

As usual the torsion term vanishes, modulo a total derivative, after the variation of spin connection $\omega$ \cite{jacobson}.

As have been shown in \cite{samuel}, the self-dual action (\ref{accionsd}) has also the interesting feature to coincide with the Ashtekar action. (\ref{accionsd}) has been also obtained in \cite{nieto1} from the Mac-Dowell-Mansouri action.

The action of Pleba\'nski is formulated in spinorial notation, obtained by
$v^i\rightarrow v_A^{\ B}=v_i\sigma_{\,A}^{i\ B}$, or $v^i=\frac{1}{2}\sigma^{i B}_A v^{\ B}_A$, where $\sigma_{\,A}^{i\ B}$ are the Pauli Matrices. For the Lorentz group the corresponding embedding is into $SL(2,C)\otimes\overline{SL(2,C)}$, given for the fundamental representation by
$v_a\rightarrow v_{A\dot B}=v_a\sigma^a_{A\dot B}$, where $\sigma_{0A\dot B}$ is the identity matrix and
$\sigma^i_{A\dot B}$ are the Pauli matrices. With these conventions, the adjoint representation of $SO(3,1)$ decomposes as,
\begin{equation}\label{descomposicion}
\sigma^a_{A\dot A}\sigma^b_{B\dot B}v_{ab}=
\epsilon_{\dot A\dot B}(\sigma^{ab}\epsilon)_{AB}v_{ab}+
\epsilon_{AB}(\epsilon\bar\sigma^{ab})_{\dot A\dot B}v_{ab}
=-\epsilon_{\dot A\dot B}v_{AB}+\epsilon_{AB}v_{\dot A\dot B},
\end{equation}
where $\sigma^{ab}=\frac{1}{4}[\sigma^a,\bar\sigma^{b}]$  and
$\bar\sigma^{ab}=\frac{1}{4}[\bar\sigma^{a},\sigma^b]$, satisfy $\epsilon^{ab}_{\ \ cd}\sigma^{cd}=2i\,\sigma^{ab}$ and $\epsilon^{ab}_{\ \ cd}\bar\sigma^{cd}=-2i\,\bar\sigma^{ab}$. Hence $v_{AB}$ and $v_{\dot A\dot B}$ are self-dual, respectively anti-self-dual.
Consistently with it, for complex $SU(2)$ vectors we get, $u^iv_i=-\frac{1}{4}u^{(+)ab}v^{(+)}_{\ ab}=
-\frac{1}{4}u^{AB}v_{AB}$.

With these elements, equation (\ref{descomp}) can be inverted to recover decomposition (\ref{descomposicion}). First we have
\begin{equation}
\Omega^{0i}=\Omega^i+\overline{\Omega^i}=
\frac{1}{2}\left(\sigma^{i B}_{A}\Omega^{\ A}_B+\overline{\sigma^{i B}_{A}\Omega^{\ A}_B}\right)
=(\sigma^{0i})^{\ B}_{A}\Omega^{\ A}_B+(\bar{\sigma}^{0i})_{\ \dot A}^{\dot B}\Omega^{\dot A}_{\ \dot B},
\end{equation}
where the bar represents the complex conjugated and,
\begin{equation}\label{reality}
\Omega^{\dot A}_{\ \dot B}=-\overline{\Omega^{\ A}_B}.
\end{equation} 
Further 
\begin{equation}
\Omega^{ij}=-\frac{i}{2}\epsilon^{ijk}\left(\sigma^{k B}_{A}\Omega^{\ A}_B-\overline{\sigma^{k B}_{A}\Omega^{\ A}_B}\right)
=(\sigma^{ij})^{\ B}_{A}\Omega^{\ A}_B+(\bar{\sigma}^{ij})_{\ \dot A}^{\dot B}\Omega^{\dot A}_{\ \dot B}.
\end{equation}
Hence
\begin{equation}\label{descomposicion1}
\Omega^{ab}=(\sigma^{ab})^{\ B}_{A}\Omega^{\ A}_B+
(\bar{\sigma}^{ab})_{\ \dot A}^{\dot B}\Omega^{\dot A}_{\ \dot B}.
\end{equation}

Further, in spinorial notation the constraints (\ref{constraints}) are
\begin{equation}
B^{AB}B_{CD}-\frac{1}{3}\delta_{(CD)}^{AB}B^{EF}B_{EF}=0
\end{equation}
whose solution is shown in \cite{Plebanski}, $B_{A}^{\ B}=\Sigma _{A}^{\ B}=\frac{1}{4}e_{A\dot A}\wedge e^{\dot AB}$. Thus considering (\ref{reality}) and (\ref{descomposicion1}), we get (\ref{solconstr})
$\Sigma^{ab}=(\sigma^{ab})^{\ B}_{A}\Sigma^{\ A}_B+
(\bar{\sigma}^{ab})_{\ \dot A}^{\dot B}\Sigma^{\dot A}_{\ \dot B}=e^a\wedge e^b.$

Pleba\'nski considers in \cite{Plebanski} the more general case in which 
$\Sigma^{\dot B}_{\ \dot A}$ and $\Sigma_A^{\ B}$ are not related by complex conjugation, see also \cite{capovilla}. Thus for each of them there are independent solutions to the constraints, called the heavenly and the hellish solutions. Gravity arises after the reality conditions are imposed, which here correspond to (\ref{reality}).
\section{Supergroup action}
The supersymmetric generalization of the Ashtekar formulation has been given and worked out by Jacobson 
\cite{jacobson}, who gave a fermionic $SL(2,C)$ formulation starting from the first order formalism. Further, Capovilla et. al. \cite{capovilla} considered fermionic constraints in addition to the constraints (\ref{constraints}). This action has been considered further in \cite{sano}, where the full supersymmetry invariance is shown, as well as the Kalb-Ramond symmetry. In \cite{ezawa} a manifest $OSp(2|1)$ invariant $BF$ action which leads to supergravity was written, whith a cosmological term given by the cosmological constant times ${\rm STr}B^2$, and in \cite{nojiri} the simplicity constraint terms, bosonic plus fermionic, were given as the supertrace under certain irreducible representations of $OSp(2|1)$. 

In this section we consider action (\ref{pleba}) generalized to $OSp(2|1)$, following the lines of the previous section. The last term in (\ref{pleba}) plays here an important role, as it avoids that the constraints eliminate too many degrees of freedom, allowing also that the consistency conditions corresponding to (\ref{consistencia}) have the right form. 

The generalization of action (\ref{pleba}) is done by making its fields transforming under the adjoint representation of $OSp(2|1)$ and substituting the trace by the supertrace. The algebra of $OSp(2|1)$ is $[t_p,t_q]=f_{pq}^{\ \ r}t_r$, where the nonvanishing components of the structure constants are,
\begin{equation}\label{algebra}
f_{ij}^{\ \ k}=-2i \epsilon_{ij}^{\ \ k},\ \
f_{iA}^{\ \ B}=\sigma_{iA}^{\ \ B}\ {\rm and} \
f_{AB}^{\ \ \ i}=\sigma^i_{AB}=\epsilon_{BC}\sigma_{A}^{i\ C}.
\end{equation}
The invariant Killing metric tensor is,
\begin{equation}        \label{eq:Killingform}
\kappa_{pq} =\left( \begin{array}{cc}
\delta_{ij} &  0    \\
0           &  \epsilon_{AB}
\end{array} \right),
\end{equation}
and STr$(T_pT_q)=2 \kappa_{pq}$.

Further, we have $B=B^{p} \, t_{p}=B^{i} \, t_{i}+B^{A} \, t_{A}$, and
$ \Omega= \Omega^{p} \, t_{p}= \Omega^{i} \, t_{i}+ \Omega^{A} \, t_{A}$  ($i=1, 2, 3$, $A=1, 2$), where $B^{A}$ and
$ \Omega^{A}$ are the fermionic differential forms corresponding to $B^i$ and $\Omega^i$.

The field strength is given by $F=d \Omega+ \Omega\wedge \Omega =F^i\, t_i+F^A\, t_A$, where,
\begin{equation}\label{f1}
F^{i}=d \Omega^{i}+i\epsilon^i_{\ jk} \Omega^j\wedge \Omega^k-\frac{1}{2} \sigma^{i\ B}_{\ A} \Omega^A\wedge \Omega_B.
\end{equation}
From which follows
\begin{equation}
F^{ab}=R^{ab}-\frac{1}{2} \sigma^{ab\ B}_{\ \ A} \Omega^A\wedge \Omega_B,
\end{equation}
where $R^{ab}$ is given by (\ref{R}). Furthermore, if we rename $ \Omega^A=\sqrt{2k\Lambda}\psi^A$, we have
\begin{equation}\label{domega}
F_{A}=\sqrt{2k\Lambda}\left[d\psi_{A}+ \Omega_{i}\wedge(\sigma^{i}\psi)_{A}\right]
=\sqrt{2k\Lambda}D\psi_A,
\end{equation}
where the covariant derivative can be written also as
$D\psi_A=d\psi_{A}-\frac{1}{2}\omega_{ab}\wedge(\sigma^{ab}\psi)_{A}$. Here $\Lambda$ denotes the square root of the cosmological constant of the preceding section, from the usual notation in supergravity.

The natural generalization of the BF part of the action of Pleba\'nski is $2i\int \frac{1}{2}{\rm STr}B \wedge F$, where the supertrace is given by \,STr$B\wedge F=2B^p\wedge F^q\,\kappa_{pq}=2B^p\wedge F_p=2(B^i\wedge F_i+B^A\wedge F_A)$. However, the straightforward generalization of the constraint term does not work as well. Indeed, it would be given by the supertraceless expression,
\begin{equation}\label{c1}
B^p B^q-\frac{1}{5}\kappa^{pq}B^r B_r=0.
\end{equation}
If we decompose these constraints, their bosonic part is given by $B^i\wedge B^j-\frac{1}{5}\delta^{ij}(B^k\wedge B_k+B^A\wedge B_A)=0$, from which we get
\begin{equation}\label{c2}
2 B^i\wedge B_i=3 B^A\wedge B_A.
\end{equation}
Substituting this equation into the preceding relation, we get the correct bosonic constraints $B^i\wedge B^j-\frac{1}{3}\delta^{ij}B^k\wedge B_k=0$, which have the solution (\ref{solconstr}). However, this solution together with (\ref{c2}) implies that the space-time volume element is nilpotent. Moreover, (\ref{c1}) leads to the constraints $B^i\wedge B^A=0$, which turn out to be too strong. Indeed, $B^i$ projects on the self-dual part and the solution is anti self-dual, i.e. $B_A=\Sigma^{(-)ab}(\sigma_a\bar\psi_b)_A$, which is not consistent with the Rarita-Schwinger action. The right fermionic constraints were given in \cite{capovilla}, and are given by $\sigma_{i(AB}B^i\wedge B_{C)}=0$. We could not find a way to overcome these drawbacks for a supergroup generalization of action (\ref{pleba0}), other than to choose action (\ref{pleba}) for the generalization. 
Hence we will consider the action
\begin{equation}
I=\frac{2i}{k}{\rm STr}\int \left[B\wedge  F+B\wedge\Phi(B)+\eta\Phi\right],
\label{plebasusy}
\end{equation}
where $\eta=\eta^pT_p$ is a four-form field in the adjoint representation, i.e.
$(T_p)_q^{\ r}=f_{qp}^{\ \ r}$. Thus
\begin{equation}
{\rm STr}\,\eta\Phi=\eta^i(2i\epsilon_i^{\ jk}\Phi_{jk}-\sigma_i^{\ AB}\Phi_{AB})
+\eta^A\sigma_A^{i\ B}(\Phi_{iB}+\Phi_{Bi}).\label{eta}
\end{equation}
The first two terms on the r.h.s, with factor $\eta^i$, mean that the antisymmetric part of $\Phi_{ij}$ and the symmetric part of $\Phi_{AB}$ vanish, with no consequences for this action. Thus in the following we will keep only the terms with $\eta^A$ in this expression.  

Further, in the same way as in the bosonic case, we have the field equations
\begin{equation}\label{eqsusy}
F_p+\Phi_p^{\ q}B_q=0,\qquad DB_p=0,
\end{equation}
from which follow the consistency conditions
\begin{equation}\label{consistenciasusy}
D\Phi_{p}^{\ q}\wedge B_q=0.
\end{equation}
Similarly to the bosonic case (\ref{consistencia3}) (see Appendix), we get
\begin{equation}\label{constraints3}
0=\epsilon^{\nu\rho\sigma\tau}B_{\nu\mu}^pD_\rho\Phi_{p}^{\ q} B_{\sigma\tau q}
=\frac{1}{4}\epsilon^{\nu\rho\sigma\tau}B_{\nu\rho}^qB_{\sigma\tau}^pD_\mu\Phi_{pq}.
\end{equation}

In order to analyze this equation, and to derive the rest of the constraints, we first decompose the matrix $\Phi^{pq}$ into its irreducible components.
We have $\Phi^{ij}=\phi^{(t)ij}+\frac{1}{3}\delta^{ij}\Phi_1$ and
$\Phi^{AB}=-\frac{1}{2}\epsilon^{AB}\Phi_2$, where $\Phi_1=\Phi^i_{\ i}$ and $\Phi_2=\Phi^A_{\ A}$. Further,
$\frac{1}{2}(\Phi^{iA}+\Phi^{Ai})=\frac{1}{8}\sigma^i_{BC}\Phi^{BCA}=\frac{1}{8}\sigma^i_{BC}(\phi^{(BCA)}+2\epsilon^{BA}\phi^C)$, where $\phi^{(BCA)}$ is fully symmetric and
$\phi_A=\frac{2}{3}\sigma_A^{i\ B}(\Phi_{iB}+\Phi_{Bi})$. 

Thus from (\ref{eta}) we get Str$\,\eta\Phi=\frac{3}{2}\eta^A\phi_A$, and the action can be written as
\begin{eqnarray}
I&=&\frac{2i}{k}\int\bigg\{2B^i\wedge F_i+2B^A\wedge F_A+\phi^{(t)ij}B_i\wedge B_j 
+\frac{1}{3}\Phi_1B^i\wedge B_i-\frac{1}{2}\Phi_2B^A\wedge B_A\nonumber\\
&&\qquad\quad+\frac{1}{4}\sigma^i_{BC}\phi^{(BCA)}B_i\wedge B_A
+\frac{1}{2}\sigma^{iAB}\phi_BB_i\wedge B_A
+\frac{3}{2}\eta^A\phi_A\bigg\}.\label{plebasusy1}
\end{eqnarray}

Hence if we variate with respect to $\phi^{(t)ij}$, we get the constraints (\ref{constraints}). Further, the variation of $\phi^{(ABC)}$ gives the fermionic constraint of Capovilla et. al. 
\cite{capovilla}
\begin{equation}
B_{(A}B_{BC)}=\sigma^i_{(AB}B_{C)}B_i=0,\label{capo}
\end{equation}
and the variation of $\eta^A$ gives the constraint
\begin{equation}\label{nuevo}
\phi_A=0.
\end{equation}

Therefore, taking into account the decomposition of $\Phi_{pq}$ and equations (\ref{constraints}), (\ref{capo}) and (\ref{nuevo}), the r.h.s. of (\ref{constraints3}) gives,
\begin{eqnarray}\label{constriccion4}
0&=&B^qB^pD_\mu\Phi_{pq}
=\left[B^iB^j\,D_\mu\Phi_{ji}
+B^iB^A\,D_\mu(\Phi_{iA}+\Phi_{Ai})+
B^AB^B\,D_\mu\Phi_{BA}\right]\nonumber\\
&=&\frac{1}{3}B^iB_{i}\,D_\mu\Phi_1+\frac{1}{2}B^AB_{A}\,D_\mu\Phi_2,
\end{eqnarray}
because
\begin{equation}
B^iB^A\,D_\mu(\Phi_{iA}+\Phi_{Ai})
=-\frac{1}{4}B_{BC}B_A\,D_\mu(\phi^{(BCA)}+2\epsilon^{BA}\phi^C).
\end{equation}
(\ref{constriccion4}) is a linear equation in which 
the coefficient of $D_\mu\Phi_1$ is bosonic, and the coefficient of $D_\mu\Phi_2$ is fermionic, thus both terms are linear independent as far as both these coefficients do not vanish. The first coefficient is the volume element and the second one does not vanish, as will be seen in the following. Therefore we have 
$D_\mu\Phi_1=D_\mu\Phi_2=0$.

Further, 
\begin{eqnarray}
D_\mu\Phi_1&=&\partial_\mu\Phi_1+2\sigma_{iAB} \Omega_\mu^A\Phi^{Bi}=\partial_\mu\Phi_1
-\frac{3}{2} \Omega_\mu^A\phi_A,\\
D_\mu\Phi_2&=&\partial_\mu\Phi_2-2\sigma_{iAB} \Omega_\mu^A\Phi^{Bi}=\partial_\mu\Phi_2
+\frac{3}{2} \Omega_\mu^A\phi_A.
\end{eqnarray}
Therefore from (\ref{nuevo}) we get $\partial_\mu\Phi_1=\partial_\mu\Phi_2=0$, and we set them to be
$\Phi_1=\Lambda^2$ and $\Phi_2=16\Lambda^2$, in such a way that the corresponding terms in 
(\ref{plebasusy1}) have the form of the volume and fermionic mass cosmological constant terms of supergravity \cite{townsend}. 

Returning to the constraints, the solution for $B^i$ is given by the self-dual part of (\ref{solconstr}), and the solution of the fermionic constraint \cite{capovilla} can be obtained making the decompostion
$B_A=-\sqrt{k}\Sigma^{ab}(\sigma_{ab}^{BC}\rho_{A(BC)}
-\bar\sigma_{ab}^{\dot B\dot C}\rho_{A(\dot B\dot C)})$, and $\rho_{A(BC)}=\rho_{(ABC)}+\frac{1}{2}(\epsilon_{AB}\rho_C+\epsilon_{AC}\rho_B)$,
from which we get from (\ref{capo}) that
$\sigma_{i(BC}B^iB_{A)}=\rho_{(ABC)}e\, d^4x$=0. Thus, the self-dual and anti-self-dual parts of $B_A$ are the irreducible components of a one-form spinor field $\bar\psi^{\,\dot A}$, which are given by $\bar\psi_{A\dot A}^{~~\,\dot A}=-8i\sqrt \Lambda\rho_A$ and
$\bar\psi_{A(\dot A\dot B)}=8i\sqrt \Lambda\rho_{A(\dot A\dot B)}$, in such a way that,
\begin{equation}
B_A=\frac{i}{4}\sqrt{\frac{k}{2\Lambda}}\Sigma^{ab}(\sigma_a\bar\psi_b)_A,
\end{equation}
from which then
\begin{equation}
B^AB_A=\frac{ik}{8\Lambda}\bar\psi_a\bar\sigma^{ab}\bar\psi_b\,e\,d^4x,
\end{equation}
which does not vanish.

Thus, the action (\ref{plebasusy1}) turns to,
\begin{equation}
I=\frac{2i}{k}\int\Big(2B^i\wedge F_i+2B^A\wedge F_A
+\frac{\Lambda^2}{3} B^i\wedge B_i-8\Lambda^2 B^A\wedge B_A\Big).\label{plebasusy3}
\end{equation}

Therefore, taking into account (\ref{f1}), (\ref{domega}) and the solution of the constraints, we get,
\begin{eqnarray}\label{SuperAccion2}
I&=&\int\Big\{-\frac{1}{2k}\left[R(\omega)+2\Lambda^2\right]
+2\Lambda\left(\psi_a\sigma^{ab}\psi_b
+\bar\psi_a\bar\sigma^{ab}\bar\psi_b\right)\Big\}e d^4x
\nonumber\\
&&\qquad+\int e^a\wedge\bar\psi_{\dot A}\wedge(\bar\sigma_aD\psi)^{\dot A}-\frac{i}{4k}\int e^a\wedge DT_a.
\end{eqnarray}

As usual, a variation of $\omega$ in this action gives,
\begin{equation}\label{SuperBF3}
\delta_\omega I=\frac{2}{k}\int e^a\wedge\left[T^b-
\frac{ik}{2}\psi^{A}\wedge(\sigma^b\bar\psi)_{A}\right]\wedge\delta\omega^{+}_{ab}=0,
\end{equation}
from which follows the supertorsion vanishing condition $T^a-\frac{ik}{2}\psi^{A}\wedge(\sigma^a\bar\psi)_{A}=0$.
From this condition and from 
$\psi^{A}\wedge(\sigma^a\bar\psi)_{A}\wedge\psi^{B}\wedge(\sigma_a\bar\psi)_{B}=0$ follow, modulo surface terms, $\int e^a\wedge DT_a=0$ and
$\int e^a\wedge \bar\psi_{\dot A}\wedge (\bar\sigma_aD\psi)^{\dot A}=\int e^a\wedge 
D\bar\psi_{\dot A}(\bar\sigma_a\psi)^{\dot A}$.
Therefore we get the supergravity action with cosmological constant of Townsend \cite{townsend},
\begin{eqnarray}\label{SuperAccion3}
I&=&\int\bigg\{-\frac{1}{2k}\left[R(\omega)+2\Lambda^2\right]
-\frac{1}{2}\epsilon^{\mu\nu\rho\sigma}\left(D_\mu\psi_\nu\sigma_\rho\bar\psi_\sigma
-\psi_\mu\sigma_\nu D_\rho\bar\psi_\sigma\right)\nonumber\\
&&\qquad+2\Lambda\big(\psi_a\sigma^{ab}\psi_b
+\bar\psi_a\bar\sigma^{ab}\bar\psi_b\big)\
\bigg\}\ e\ d^4x,
\end{eqnarray}

Note that the values of the parameters in this action have been chosen in such a way that it is consistent, i.e. that the imaginary part of the final action does not generate new equations of motion, and that the parameters have a physical interpretation. In particular STr$\Phi=\phi_1+\phi_2$ is invariant and constant, so we expect naturally that to it corresponds only one scale, given by the cosmological constant.

The transformations corresponding to (\ref{kr1})-(\ref{kr3}) under which action (\ref{plebasusy}) is invariant are easily obtained, 
\begin{eqnarray}
\delta_v B_{\mu\nu}^p&=&-D_\mu(v^\rho B_{\nu\rho}^{\ \ p})+D_\nu(v^\rho B_{\mu\rho}^{\ \ p}),\label{skr1}\\
\delta_v \Omega_\mu^p&=&2v^\nu\Phi^{pq}B_{\mu\nu}^{\ \ q},\label{skr2}\\ 
\delta_v\Phi^{pq}&=&v^\mu D_\mu\Phi^{pq},\label{skr3}\\
\delta_v\eta^p&=&D_\mu(v^\mu\eta^p).\label{skr4}
\end{eqnarray}
Similarly to the bosonic case, the on-shell application of these transformations corresponds to diffeomorphisms plus field dependent gauge transformations, giving the same equations as in (\ref{diff1})-(\ref{diff4}), where now the gauge transformations correspond to $OSp(2|1)$.

\section{Conclusions}
The BF formulation of gravity is an interesting field of study \cite{KrasnovR,KrasnovSF,KrasnovC,KrasnovD,smolin1,smolin2}. Its group structure makes it suitable for the study of quantum gravity. One important aspect of it is given by its constraints.
In this work we consider its extension to supergravity, first considering that a straightforward supergroup generalization of the action of Pleba\'nski is inconsistent, because the constraints lead to a nilpotent volume element. This fact motivated us, in the bosonic case, to consider a generalization (\ref{pleba}) of the action given by Krasnov \cite{KrasnovD}, which has a sort of duality to Pleba\'nski action, because it turns out after application of the consistency conditions of the equations of motion and the constraints, which require that this field $\phi$ is constant. 
Further we considered the $OSp(2|1)$ extension of this action (\ref{plebasusy}), with an additional constraint which contributes at various stages of the calculations of the fermionic constraints, and of the cosmological constant terms. Similar to the bosonic case, the consistency conditions of the equations of motion, the bosonic and fermionic simplicity constraints, and the additional constraint, lead to an action which in this case is the action of supergravity with cosmological constant given by Townsend \cite{townsend}. Both actions, the bosonic and the supergroup ones, have a non-linear Kalb-Ramond vector symmetry, which on-shell corresponds to diffeomorphism invariance.

The BF action of gravity has been considered as the starting point for various generalizations as well as quantization proposals for gravity. It would be interesting to consider the quantization of the approach presented in this work, in particular for spin-foam models. Similarly for the generalizations proposed in \cite{montesinos,smolin2,KrasnovR,KrasnovSF,KrasnovC,KrasnovHD,KrasnovD}.

Also the introduction of matter into the action \cite{capovilla,Krasnov} or the generalization to higher dimensions \cite{KrasnovHD} could be considered. The consequences of the presence of a cosmological constant regarding the deformation of the symmetry group of discretized models could be also studied \cite{baez,baez1}, as well as the consequences of degenerated metrics \cite{reisenberger2}.

\appendix
\section{}
In this appendix we give an algebraic version and a slight generalization of the identity for two forms given in \cite{KrasnovD}, $\mathnormal{i}_\xi B^{(i}\wedge B^{j)}=\frac{1}{2}\mathnormal{i}_\xi(B^{(i}\wedge B^{j)})$. 
Let us consider the quantity
\begin{equation}\label{a1}
M_{a[b}M_{cd]}=\frac{1}{3}(M_{ab}M_{cd}+M_{ac}M_{db}+M_{ad}M_{bc}).
\end{equation}

It is easy to show that,
\begin{equation}
M_{a[b}M_{cd]}=-M_{b[c}M_{da]}=M_{c[d}M_{ab]}=-M_{d[a}M_{bc]}.
\end{equation}
Therefore,
\begin{equation}
M_{a[b}M_{cd]}=M_{[ab}M_{cd]}.
\end{equation}

All these steps can be repeated if we have a bilinear form with a
symmetric $\Phi_{ij}$, and write instead of (\ref{a1})
\begin{equation}\label{a2}
\Phi_{ij}M^i_{a[b}M^j_{cd]}=\Phi_{ij}M^i_{[ab}M^j_{cd]}.
\end{equation}
or with an (anti)symmetric $\Phi_{pq}$,
\begin{equation}\label{a3}
\Phi_{pq}M^q_{a[b}M^p_{cd]}=\Phi_{pq}M^q_{[ab}M^p_{cd]}.
\end{equation}

Thus, in general (in four dimensions, considering a Minkowski signature)
\begin{equation}
\Phi_{pq}M^q_{a[b}M^p_{cd]}
=-\frac{1}{24}\epsilon_{abcd}\epsilon^{efgh}\Phi_{pq}M^q_{ef}M^p_{gh}.
\end{equation}
In particular we have,
\begin{equation}\label{identidad}
\epsilon^{\nu\rho\sigma\tau}D_\nu\Phi_{pq}B^q_{\mu[\rho}B^p_{\sigma\tau]}
=\frac{1}{4}\epsilon^{\nu\rho\sigma\tau}D_\mu\Phi_{pq}B^q_{\nu\rho}B^p_{\sigma\tau}.
\end{equation}

This would work as well with anyonic quantities.

\section*{Acknowledgments}
We thank O. Obreg\'on, H. Garc\'\i a Compe\'an  and G. Garc\'\i a for useful discussions.
This work was supported in part by CONACyT M\'exico Grant 51306 and by BUAP-VIEP grants.

\end{document}